\documentclass[onecolumn,floatfix,superscriptaddress,a4paper,
               showpacs,showkeys,nofootinbib,preprint]{revtex4}
\usepackage{epsfig}
\usepackage{latexsym}
\usepackage{xspace}
\usepackage{hyperref}
\usepackage[latin2]{inputenc}
\usepackage{indentfirst}
\usepackage{enumerate}

\usepackage{amsmath}
\usepackage{amssymb}
\usepackage[english]{babel}
\usepackage{url}

\begin{document}

\title{Mean transverse mass
of hadrons \\
in proton-proton reactions }
\author{V. Yu. Vovchenko}
\affiliation{
Taras Shevchenko National University of Kiev, Kiev 03022, Ukraine}
\affiliation{
Frankfurt Institute for Advanced Studies, Frankfurt 60438, Germany}
\affiliation{
Goethe University, Frankfurt 60325, Germany}
\author{D. V. Anchishkin}
\affiliation{
Bogolyubov Institute for Theoretical Physics, Kiev 03680, Ukraine}
\affiliation{
Taras Shevchenko National University of Kiev, Kiev 03022, Ukraine}
\author{M. I. Gorenstein}
\affiliation{
Bogolyubov Institute for Theoretical Physics, Kiev 03680, Ukraine}
\affiliation{
Frankfurt Institute for Advanced Studies, Frankfurt 60438, Germany}

\date{\today}



\begin{abstract}
An energy dependence of the mean transverse mass $\langle m_T\rangle$ at
mid-rapidity in  proton-proton ($p+p$) reactions is studied within the
ultra-relativistic quantum molecular dynamics (UrQMD).
The UrQMD model predicts a nonmonotonous dependence of $\langle m_T\rangle$ on
collision energy for several hadron species:
for $\pi^+$, $p$, $K^+$, and $\Lambda$ the mean transverse mass has a maximum
at the center of mass energy region  $5\le \sqrt{s}\le 8$~GeV.
These results
are a consequence of an interplay of two contributions:
1) excitations and decays of the baryonic resonances $N^*$ and $\Delta$;
2) excitations and decays of the baryonic strings.
The UrQMD results do not show any nonmonotonous dependence of
$\langle m_T\rangle$ on $\sqrt{s}$ for $\pi^-$, $K^{-}$, and antiprotons.
Whether a nonmonotonous dependence of $\langle m_T\rangle$ at mid-rapidity on
the collision energy for $\pi^+$,  $p$, $K^+$, and $\Lambda$ is relevant for
real $p+p$ interactions will be soon checked experimentally by the
NA61/SHINE Collaboration.
\end{abstract}

\maketitle

Experimental data on hadron production in central Pb+Pb collisions
obtained by the NA49 Collaboration \cite{NA49-1,NA49-3,NA49-2}
at the Super Proton Synchrotron (SPS) of the European Organization for Nuclear
Research (CERN) are consistent with the onset of deconfinement in central
nucleus-nucleus collisions at about $\sqrt{s_{NN}}=(7-8)$~GeV \cite{GaGo,GGS1,GGS2},
where $\sqrt{s_{NN}}$ is the center of mass energy of two nucleons from colliding
nuclei.
An important experimental result in this context is an observation
of a {\it step}-like behavior of a mean value of the transverse mass
$m_T=\sqrt{m^2+p_T^2}$, where $m$ is a particle mass and $p_T$ its transverse momentum.
The mean transverse mass $\langle m_T\rangle$ for
different hadron species ($\pi^-$, $K^{\pm}$, $p$, and $\overline{p}$) measured
at mid-rapidity in central Pb+Pb collisions demonstrates
an increase at small and large collision energies,
but it remains approximately constant in the SPS energy range $\sqrt{s_{NN}}=7.6-17.3$~GeV
(see Fig.~8 in Ref.~\cite{NA49-2}). This is the energy region, where one expects the transition between
confined and deconfined matter with the creation of mixed phase.
Such an energy dependence of $\langle m_T\rangle$ is indeed typical for a
$1^{st}$ order phase transition in the mixed-phase region~\cite{vHove,Shuryak,GGB:2003}.
In existing up to now data,
there were no indications on such type of $\langle m_T\rangle$ dependence on the
center of mass energy $\sqrt{s}$
in proton-proton ($p+p$) reactions.

A further progress in understanding the effects related to the onset of
deconfinement can be achieved by a new comprehensive study of hadron production
in $p+p$, proton-nucleus, and nucleus-nucleus collisions.
This  motivated the present NA61/SHINE ion programme at the SPS CERN devoted to
the system size and energy scan~\cite{Ga:2009,NA61facility} and the Beam
Energy Scan (BES) programme at the Relativistic Heavy Ion Collider (RHIC)
of the Brookhaven National Laboratory (BNL)~\cite{RHIC}.
These efforts will be extended by the future
Compressed Baryonic Matter (CBM) experiment at
the Facility for Antiproton and Ion Research (FAIR)~\cite{CBMPhysicsBook,cbm},
which will employ high-luminosity beams and large-acceptance detectors to
study system-size and energy dependence of hadron production in proton-nucleus and nucleus-nucleus collisions.

\begin{figure}[ht]
\centering
\includegraphics[width=0.49\textwidth]{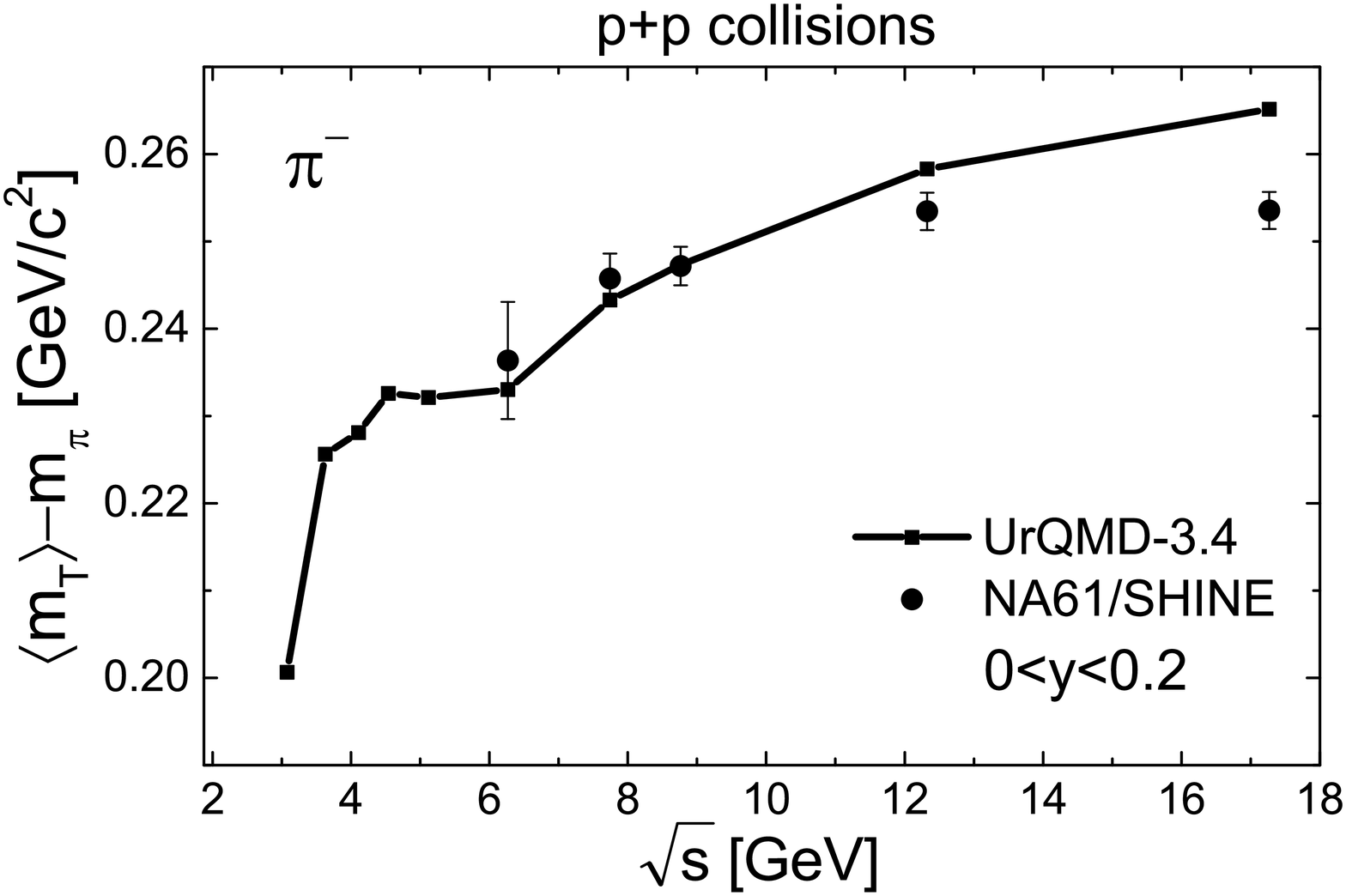}
\includegraphics[width=0.49\textwidth]{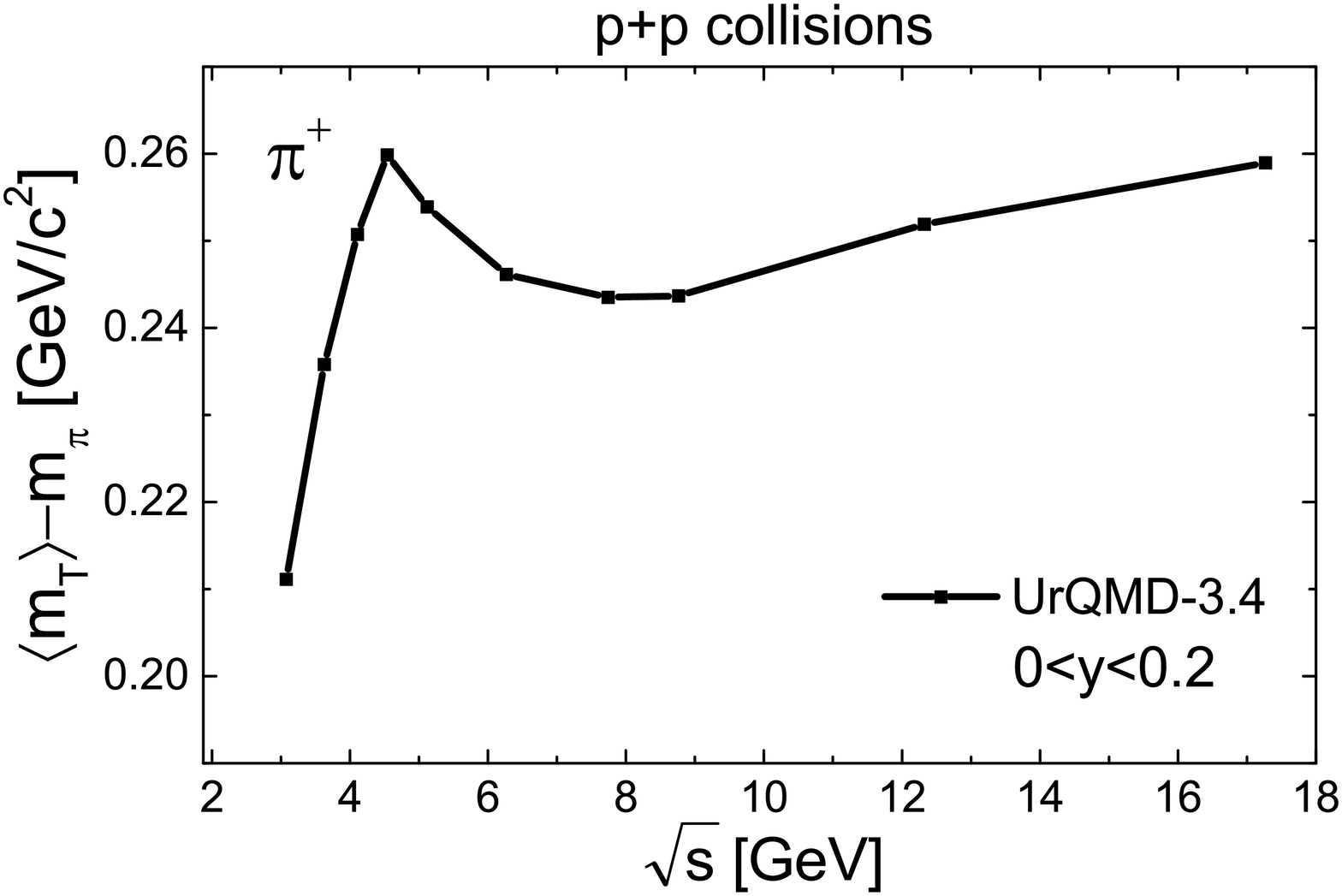}
\caption[]{The mean transverse mass $\langle m_T\rangle - m_\pi$ of
$\pi^-$ ({\it left} panel) and $\pi^+$ ({\it right} panel)
in $p+p$ reactions as a function of the center of mass energy $\sqrt{s}$. The rapidity window
in the center of mass system is $0<y<0.2$.
The solid lines show the results of UrQMD-3.4 calculations.
The symbols with error bars in the {\it left} panel are the data from Ref.~\cite{NA61pp}. }
\label{fig-pi}
\end{figure}

In Ref.~\cite{VAG}, we analyzed  recent NA61/SHINE data \cite{NA61pp}
on $\pi^-$ spectra in $p+p$ reactions within the ultra-relativistic quantum
molecular dynamics (UrQMD) model~\cite{UrQMD1998,UrQMD1999}. The version UrQMD-3.3p2 \cite{UrQMD:2008}
was used in Ref.~\cite{VAG}.
Recently a new version, UrQMD-3.4 \cite{UrQMD:2014}, was released.
It will be used in the present study.
Both the data and the UrQMD simulations demonstrate a monotonous increase
of $\langle m_T\rangle$ with  collision energy for $\pi^-$ at mid-rapidity.
This is shown in Fig.~\ref{fig-pi} {\it left}.
Unexpectedly, the UrQMD has predicted a nonmonotonous dependence of
$\langle m_T\rangle$ on collision energy for $\pi^+$ in $p+p$ reactions:
the mean transverse mass of positively charged pions evaluated at mid-rapidity
has a maximum at $\sqrt{s}\cong 4.5$~GeV and then decreases notably with the collision energy inside the
region of $\sqrt{s}=5 - 8$~GeV. This is shown in Fig.~\ref{fig-pi} {\it right}.

The nonmonotonous dependence of $\langle m_T\rangle$ on $\sqrt{s}$
seen in Fig.~\ref{fig-pi} {\it right} for $\pi^+$ accepted at mid-rapidity
are absent for the UrQMD results taken at all rapidities.
This is in an agreement with a compilation of the
old $p+p$ data in Ref.~\cite{compil}.

A physical origin of the nonmonotonous  dependence of $\langle m_T\rangle$
on $\sqrt{s}$ in the UrQMD model
for  $\pi^+$ in $p+p$ reactions is connected to a
presence of two different sources of pions.
Main inelastic reactions at small energies are the following \cite{UrQMD1998,UrQMD1999}:
\begin{eqnarray}
&& p+p\rightarrow p+\Delta^+~,~~~~~p+p\rightarrow n+\Delta^{++}~,~~~~~
p+p\rightarrow \Delta^++\Delta^+~,~~~~~ p+p\rightarrow \Delta^0+\Delta^{++}~,
\nonumber \\
&& p+p\rightarrow p+N^+~,~~~~~ p+p\rightarrow N^++\Delta^+~,~~~~~p+p\rightarrow N^0+\Delta^{++}~.
\label{pppi}
\end{eqnarray}
In the UrQMD transport model, the states of $N^*$ with $m=1440,\ldots, 2250$~MeV and
the states of $\Delta$ with $m=1232,\ldots,1950$~MeV are included.
Specific charge states of $N^*$ and $\Delta$ emerged in Eq.~(\ref{pppi})
explain a difference in a production
of $\pi^-$ and $\pi^+$:
only $\Delta^0$ and $N^0$ may produce $\pi^-$ after their decays,
whereas $\Delta^+$ and $N^+$ may produce only $\pi^+$,
and $\Delta^{++}$ has to produce  $\pi^+$.

The reactions listed in (\ref{pppi}) give the dominant contribution to the $p+p$ inelastic
cross section at small collision energies.
However, at $\sqrt{s}\ge 4$~GeV  the excitations of baryonic strings,
\begin{equation}
p~+~p~\rightarrow ~{\rm String}~+~ {\rm String}~,
\label{string}
\end{equation}
open the new channels of hadron production. At collision energies
$\sqrt{s} > 6$~GeV the string production dominates in the UrQMD
description of inelastic $p+p$ cross section \cite{UrQMD1998}.

We can distinguish two sources contributing to hadron production in $p+p$ reactions:
1)~the excitations of baryonic resonances listed in (\ref{pppi}) and
their decays to final hadrons;~ 2) the excitation of strings
according to (\ref{string}) and their decays to final hadrons.
The mean transverse mass of a final hadron can be presented
as
\begin{equation}
\langle m_T\rangle~=~ f_B \, \langle m_T\rangle_B~+~f_S\, \langle m_T\rangle_S~,
\label{partial}
\end{equation}
where $f_B(\sqrt{s})$ and $f_S(\sqrt{s})$ are the
fractions of  multiplicities for a given hadron species from excited baryons (\ref{pppi}) and
from strings (\ref{string}), respectively. In Eq.~(\ref{partial}),
$\langle m_T\rangle_B$ and $ \langle m_T\rangle_S$, also dependent on
$\sqrt{s}$, are the values of the mean transverse mass
for a hadron originating from these two sources.
At SPS energies, the UrQMD gives $f_S>f_B$ for $\pi^+$. Besides, $f_S$ increases and $f_B$
decreases with $\sqrt{s}$.
On the other hand, there is an opposite inequality,
$\langle m_T\rangle_B > \langle m_T\rangle_S $, between the 
transverse masses of $\pi^+$ produced by baryons and by strings. This has
a simple kinematic origin. To produce $\pi^+$ at mid-rapidity according to
Eq.~(\ref{pppi}) baryonic resonances $N^*$ or $\Delta$
should be themselves created in the mid-rapidity region. Thus, with increasing $\sqrt{s}$
transverse momentum and/or mass of
the baryonic resonance has to increase. This leads to an increase
of $\langle m_T\rangle_B$ of final $\pi^+$ with $\sqrt{s}$.
An interplay of these two contributions to $\langle m_T\rangle$ in
Eq.~(\ref{partial}) leads to a nonmonotonous dependence on $\sqrt{s}$
of the
resulting mean transverse mass
as it seen in Fig.~\ref{fig-pi} {\it right} (see Ref.~\cite{VAG} for further details).

\begin{figure}[ht]
\centering
\includegraphics[width=0.49\textwidth]{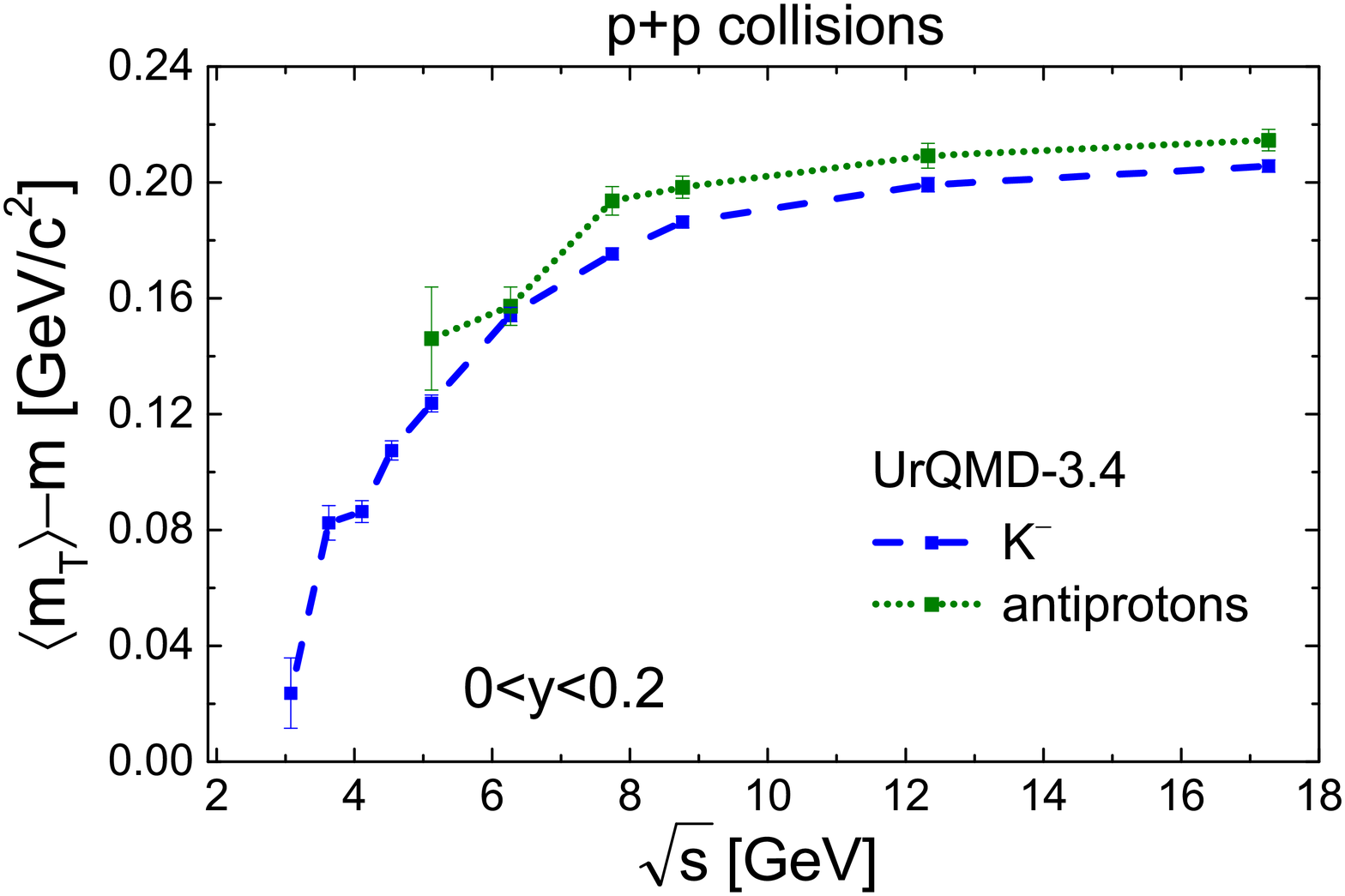}
\includegraphics[width=0.49\textwidth]{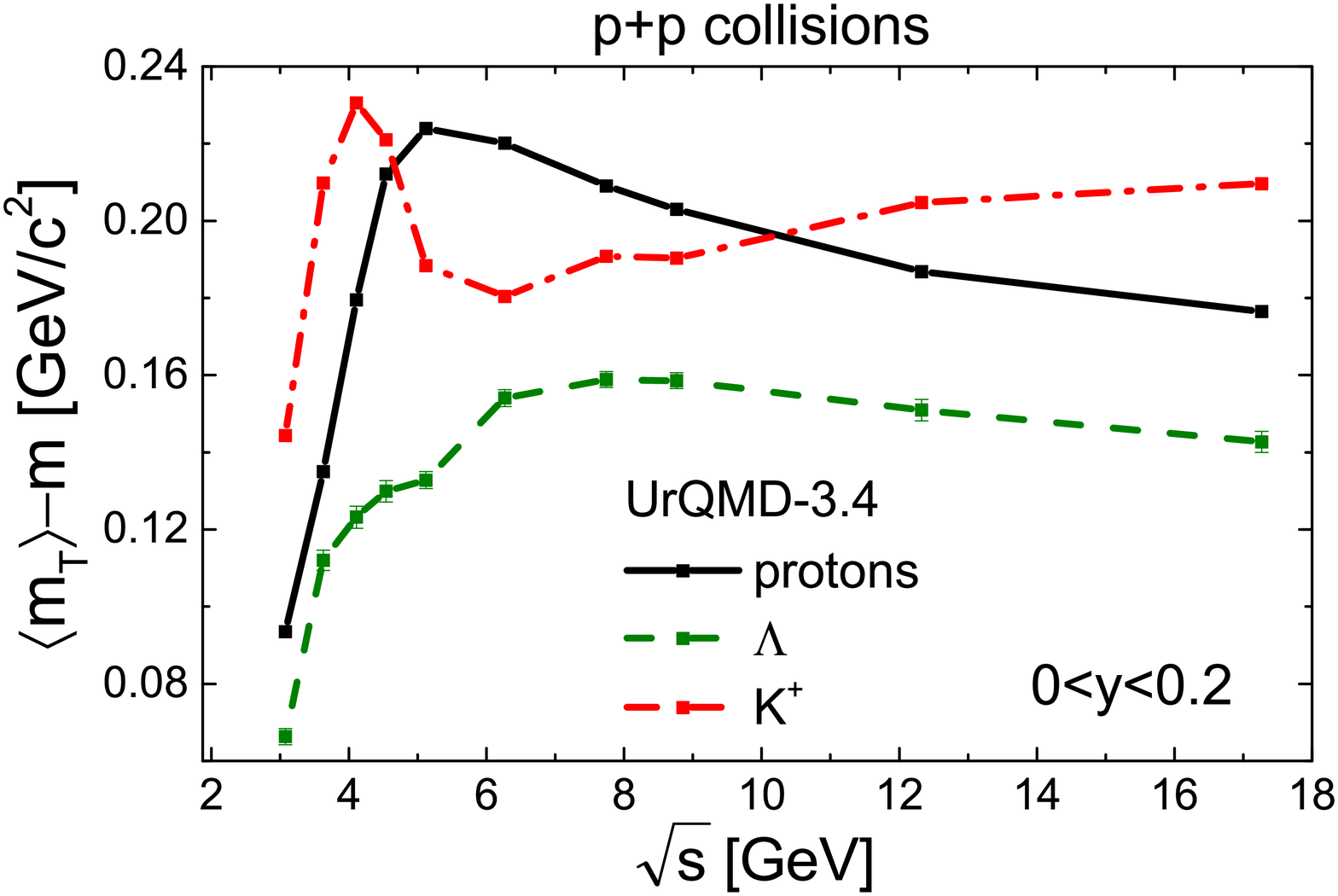}

\caption[]{The results of the UrQMD-3.4 calculations for the mean transverse
masses $\langle m_T\rangle - m$ in $p+p$ reactions as a function of collision energy.
The rapidity window in the center of mass system is $0<y<0.2$.
The error bars depict statistical uncertainties.
{\it Left}: $K^-$ (dashed line) and anti-protons (dotted line).
{\it Right:} Protons (solid line), $\Lambda$ (dashed line), and
 $K^+$ (dashed-dotted line).
}
\label{fig-p}
\end{figure}

Note that $K^-$ and $\overline{p}$ can not appear from decays of
the baryonic resonances contributed to Eq.~(\ref{pppi}).
In Fig.~\ref{fig-p} {\it left} we present the UrQMD values of $\langle m_T\rangle - m$
at mid-rapidity for these hadrons
as the functions of collision energy. No nonmonotonous dependence of $\langle m_T\rangle$
on $\sqrt{s}$ are seen.

Protons are obviously produced from decays
of baryonic resonances excited according to Eq.~(\ref{pppi}).
For heavy $N^*$ the decay channel $N^*\rightarrow \Lambda +K^+$ is included
in the UrQMD model.
Therefore, Eq.~(\ref{partial}) can be also applied to a description of $\langle m_T\rangle$
for  $p$, $\Lambda$, and $K^+$ in  $p+p$ reactions. These UrQMD
results are shown in Fig.~\ref{fig-p} {\it right}.
A nonmonotonous dependence of $\langle m_T\rangle -m$ on $\sqrt{s}$ at
mid-rapidity ($0<y<0.2$) are observed for these hadron species.
Like in a case of $\pi^+$, it
disappears if final hadrons are detected in unrestricted rapidity window.

\begin{figure}[ht]
\centering
\includegraphics[width=0.49\textwidth]{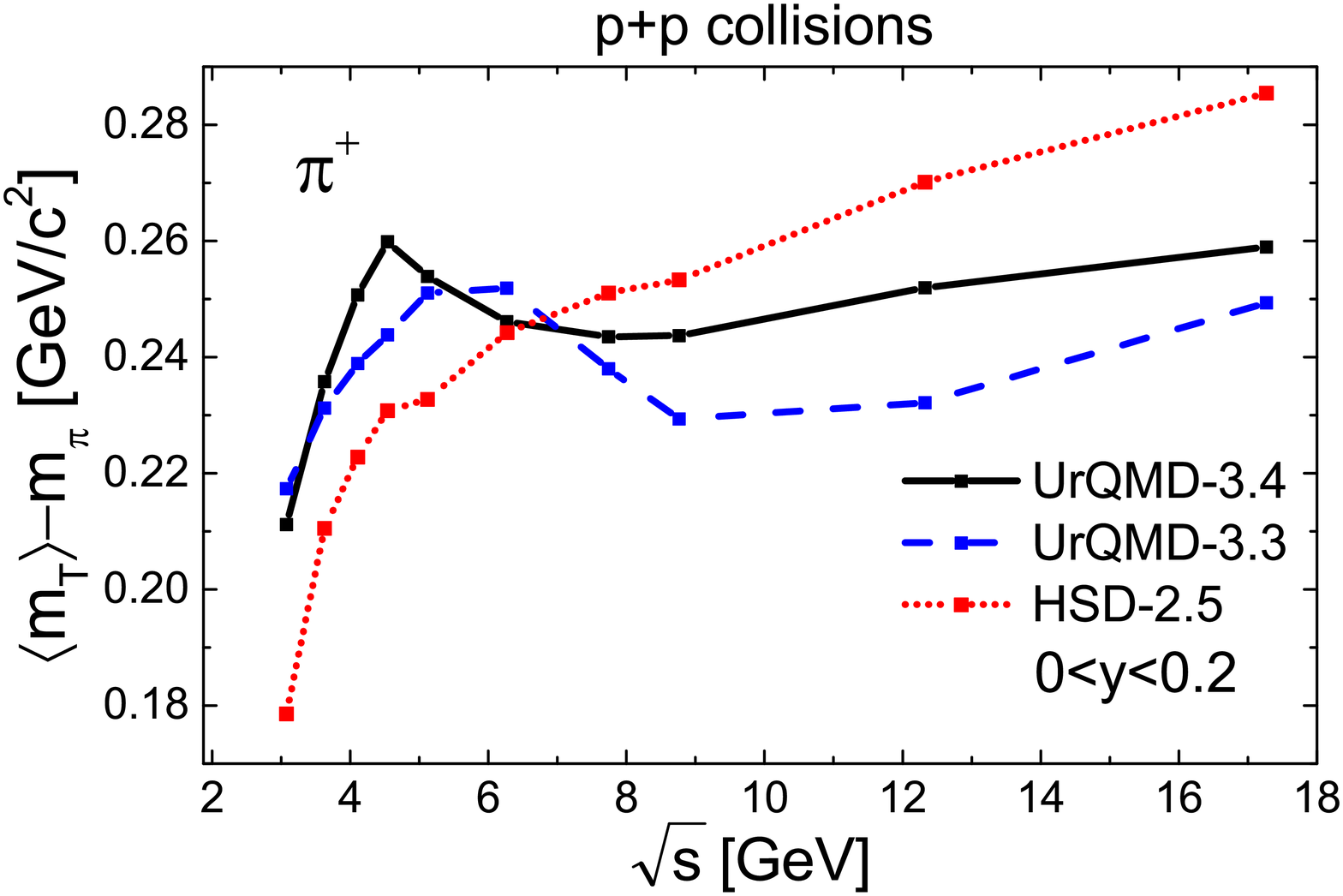}
\includegraphics[width=0.49\textwidth]{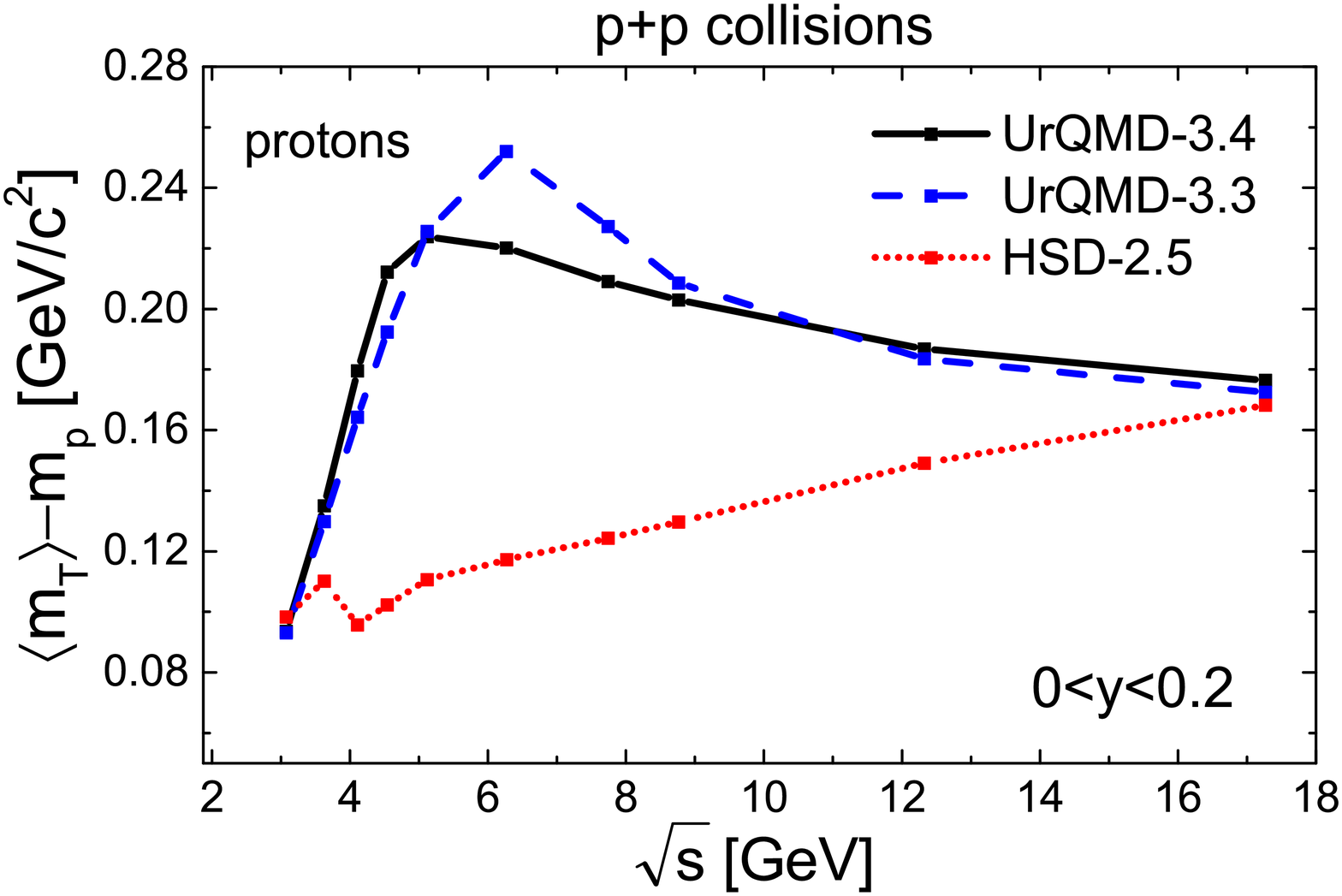}
\caption[]{The mean transverse mass $\langle m_T\rangle - m$
of $\pi^+$ ({\it left} panel) and protons ({\it right} panel)
in $p+p$ reactions as a function of collision energy.
The rapidity window in the center of mass system is $0<y<0.2$.
Solid lines correspond to the results of UrQMD-3.4, dashed ones to
UrQMD-3.3, and dotted ones to HSD-2.5.
}
\label{fig-HSD}
\end{figure}
A nonmonotonous energy dependence for the mid-rapidity values
of $\langle m_T\rangle$ for $\pi^+$,  $p$, $K^+$, and $\Lambda$ in
$p+p$ reactions observed in the UrQMD  calculations is a consequence
of an interplay of the contributions to the particle production  from  decays of the baryonic resonances and
strings.
Note that UrQMD  includes all  $N^*$ and $\Delta$ states
listed in the Particle Data Tables. A main difference between the
baryonic resonances and strings consists of in rather different
physical formulation of their decays to final hadrons.
Note, however, that experimental information about the decay branching
ratios of the high-mass resonance states is rather poor.
In Figs.~\ref{fig-HSD} {\it left} and {\it right} we present the results
for $\pi^+$ and protons, respectively,
obtained within two different versions of UrQMD, 3.3 and 3.4, as well as
within the Hadron String Dynamics (HSD-2.5) \cite{HSD1,HSD2,HSD3}.
The number of baryon species in the HSD model is smaller than that in UrQMD.
In fact, not all baryonic resonances from the Particle Data Tables are included.
Note that inelastic $p+p$ reactions above $\sqrt{s} \cong 2.6$~GeV
are described in HSD-2.5 by the FRITIOF string model~\cite{FRITIOF}.

As seen from Fig.~\ref{fig-HSD}, both UrQMD-3.3 and UrQMD-3.4 predict a
nonmonotonous dependence of $\langle m_T\rangle$ on $\sqrt{s}$ for $\pi^+$ and protons at
mid-rapidity, whereas the HSD results give a monotonous increase of
$\langle m_T\rangle$ with collision energy for both $\pi^+$ and protons.

In conclusion, within the UrQMD simulations we have observed a nonmonotonous
dependence on the collision energy of the mean transverse mass $\langle m_T\rangle$ at mid-rapidity
in inelastic $p+p$ collisions: a drop
for $\pi^+$ at  $5\le \sqrt{s}\le 8$~GeV,
a drop for $K^+$ at $4 \le \sqrt{s}\le 7$~GeV,
a decrease for protons at $5 \le \sqrt{s}\le 17.3$~GeV
and a decrease for $\Lambda$ at $8\le \sqrt{s}\le 17.3$~GeV.
These results are the specific feature of UrQMD due to a presence of two mechanisms of
  the particle production: the decays of baryonic resonances, $N^*$ and $\Delta$,  and strings.
 An interplay of these sources of particles at different collision energies results in nonmonotonous
 dependence of $\langle m_T \rangle$ on $\sqrt{s}$.
Such a nonmonotonous dependence is not found
in the HSD model, where the mass spectrum of $N^*$ and $\Delta$ is underpopulated.
Whether a nonmonotonous dependence
of $\langle m_T\rangle$ on $\sqrt{s}$ at mid-rapidity for $\pi^+$, protons, $K^+$, and $\Lambda$ is
relevant for real $p+p$ interactions
will be checked soon experimentally by the NA61/SHINE Collaboration.
These experimental data can be used for evaluation of the unknown
decay parameters of heavy resonances and inclusion of this information in the transport models.

\begin{acknowledgments}
%
We would like to thank Elena Bratkovskaya
and Marek Ga\'zdzicki
for fruitful discussions and comments.
Publication is based on the research provided by the grant support of the State 
Fund for Fundamental Research (project No. F58/175-2014). The work was partially supported 
by the Helmholtz International Center for FAIR within the LOEWE program of the State of Hesse and by the Program of 
Fundamental Research of the Department of Physics and Astronomy of NAS of Ukraine
(project No. 0112U000056).
\end{acknowledgments}


\end{document}